\documentclass[utf8]{Preprint}

\usepackage{url,hyperref,lineno,microtype,subcaption}
\usepackage{mathabx}
\usepackage{cleveref}
\usepackage[onehalfspacing]{setspace}
\usepackage{booktabs}
\usepackage{multirow}
\usepackage{xcolor}
\definecolor{goodgreen}{RGB}{33, 177, 35}
\definecolor{badred}{RGB}{186, 40, 29}

\def\keyFont{\fontsize{8}{11}\helveticabold }

\def\firstAuthorLast{Frattini {et~al.}}

\def\Authors{Julian Frattini\,$^{1,*}$, Jannik Fischbach\,$^{2,3}$, Davide Fucci\,$^{4}$, Michael Unterkalmsteiner\,$^{4}$, Daniel Mendez\,$^{3,4}$, Robert Feldt\,$^{1,5}$, Richard Torkar\,$^{1,6}$}
\def\Address{
    $^{1}$Department of Computer Science and Engineering, Chalmers University of Technology and Gothenburg University, Gothenburg, Sweden \\
    $^{2}$Netlight Consulting GmbH, Munich, Germany \\
    $^{3}$fortiss GmbH, Munich, Germany \\
    $^{4}$Software Engineering Research Lab, Blekinge Institute of Technology, Karlskrona, Sweden \\
    $^{5}$Mid Sweden University, Östersund, Sweden \\
    $^{6}$Stellenbosch Institute for Advanced Study (STIAS), Stellenbosch, South Africa
    }
    
\def\corrAuthor{Julian Frattini}
\def\corrEmail{julian.frattini@chalmers.se}

\begin{document}
\onecolumn
\firstpage{1}

%\title[Managing Variance Theories]{Replications, Revisions, and Reanalyses: Managing Variance Theories in Software Engineering} 
\title[Managing Empirical Evidence]{Replications, Revisions, and Reanalyses: Managing Empirical Evidence in Software Engineering} 

% dummy fields to be kept this way
\author[\firstAuthorLast ]{\Authors} 
\address{} 
\correspondance{} 
\extraAuth{}
\maketitle

\begin{abstract}
\textbf{Background:}
One aspired outcome of empirical research on quantitative data is a \emph{variance theory}, i.e., a quantification of the effect of an independent on a dependent variables. 
In software engineering (SE) research, variance theories quantify--—among others—--the impact of tools, techniques, and other treatments on software development outcomes like productivity, cost-efficiency, and defect-detection rates. 
The validity of variance theories stems from the \emph{synthesis} of multiple pieces of evidence, which increases its validity beyond the findings of a single study. 
\textbf{Gap:}
However, research synthesis in SE is rare and---if done---mostly limited to purely narrative syntheses.
At best, researchers perform meta-analyses to synthesize variance theories from several quantitative results.
But even meta-analyses only produce reliable results when synthesizing exact replications yet fail to generalize from variations. 
\textbf{Goal:}
We aim to extend the frontier of research synthesis beyond the state-of-the-art to systematically manage empirical evidence and its evolution.
\textbf{Method:} 
We apply method engineering to construct a framework for research synthesis from proven, individual method fragments.
The framework allows researchers to put new evidence in a clear relation to an existing body of evidence and systematically expand knowledge about a studied phenomenon. 
We demonstrate the application of this framework to two fields of research by explicitly modeling the relationship between existing pieces of evidence. 
\textbf{Result:}
The framework puts three types of evolution of evidence into relation:
(1)~\emph{replications} investigate the same hypothesis in a new context to improve external validity,
(2)~\emph{revisions} challenge an existing hypothesis to improve internal validity, and
(3)~\emph{reanalyses} replace analysis methods to improve conclusion validity.
Through a systematic evolution of evidence and clear assessment criteria for each dimension of validity, the proposed framework can determine the frontier of a field of research.
\textbf{Conclusion:}
The framework provides a perspective to systematically evolve empirical evidence in SE, supporting more constructive and productive advances in our field.

\tiny
 \keyFont{ \section{Keywords:} Research synthesis, causal inference, replication, variance theory, theory evolution}
\end{abstract}

\section{Introduction}
\label{sec:intro}

Software engineering (SE) research ultimately aims to benefit and support SE practitioners~\cite{kitchenham2004evidence,budgen2013case}, partially by proposing and evaluating new tools, techniques, and approaches.
This process of \emph{knowledge translation} from research to practice requires both \emph{knowledge creation} (i.e., gathering empirical evidence in primary studies) and \emph{knowledge application} (i.e., providing this evidence to practitioners~\cite{santos2020research}).
However, individual primary studies do not provide convincing decision support to practitioners on their own due to their limited scope and context~\cite{lo2015practitioners,devanbu2016belief,franch2020practitioners,juristo2011role}.
Hence, a necessary step between the creation and application of knowledge is the \emph{synthesis} of primary studies.

One significant product of synthesizing quantitative research is a \emph{variance theory}.
Variance theories estimate the variance of a dependent variable caused by one or more independent variables~\cite{ralph2018toward}, i.e., the effect of independent on dependent variables.
In SE research, variance theories provide decision support by quantifying the strength of the effect of, among others, tools, technologies, and human factors.
For example, the synthesis of 27 primary studies about the effect of test-driven development (TDD) on code quality and developer productivity by Rafique and Mi{\v{s}}i{\'c} determined that ``TDD has a small positive effect on quality but little to no discernible effect on productivity''~\cite{rafique2012effects}.

However, research synthesis is rarely practiced in SE and, when done, mostly limited to narrative syntheses~\cite{santos2020research}.
Only in very rare cases does SE research perform a \emph{meta-analysis} which synthesizes multiple pieces of quantitative evidence~\cite{santos2020research}.
Yet, even though certain forms of meta-analysis excel at synthesizing evidence from quantitative studies, they only produce usable results under certain conditions~\cite{pickard1998combining}.
For example, if individual pieces of evidence come to contradictory conclusions, a meta-analysis would be inconclusive and the resulting variance theory could not offer decisive decision support~\cite{hayes1999research}.
In addition, current research synthesis practices struggle to accommodate more complex relationships between individual pieces of evidence:
Interaction or mediation effects can be identified post-hoc only if candidate mediators are consistently operationalized and recorded~\cite{rafique2012effects,yang2002meta}, and many meta-analysis methods invite the unreflected usage of statistical techniques~\cite{santos2020research,kitchenham2020meta}.
As a consequence, resulting variance theories can be invalid and even detrimental when practitioners use them for their decisions.

In this work, we aim to extend the frontier of current research synthesis practices in SE beyond traditional meta-analyses.
We apply method engineering to assemble a conceptual framework from several fragments of meta-research that address the current shortcomings in research synthesis practices (\Cref{sec:framework}).
We formally define quantitative, empirical evidence (\Cref{sec:framework:evidence}), how two pieces of evidence can relate to each other (\Cref{sec:framework:relationship}), and derive how variance theories emerge and evolve systematically (\Cref{sec:framework:evolution}).
Finally, we demonstrate the framework by applying it to active fields of SE research (\Cref{sec:application}).
After discussing our results (\Cref{sec:discussion}) we conclude (\Cref{sec:conclusion}) with the aim to extend the perspective on research synthesis to obtain more valid variance theories from SE research.
%This shows how it can guide researchers toward research agendas that are more \emph{coherent}, since relationships between pieces of evidence become explicit, and more \emph{productive}, since new contributions can improve upon the revealed frontier of a research field.
While a useful method for synthesizing existing evidence, the proposed framework holds particular value to guide researchers toward more coherent and productive research agendas in the future.

\section{Related Work}
\label{sec:related}

Research synthesis (\Cref{sec:related:synthesis}) is a necessary step in empirical software engineering to produce variance theories (\Cref{sec:related:theory}), a valid and reliable estimation of how variables affect other variables.
Variance theories aim to have greater \emph{validity} than single pieces of evidence yet often still fail to achieve it (\Cref{sec:related:validity}) which motivates our approach.

\subsection{Research Synthesis}
\label{sec:related:synthesis}

The purpose of any endeavor in SE research is to support SE practitioners~\cite{budgen2013case}.
This requires translating knowledge created in research to practice~\cite{beecham2014making,cartaxo2016evidence}.
However, previous research has shown that individual empirical studies do not provide convincing evidence to practitioners~\cite{lo2015practitioners}.
A single study cannot compete with the beliefs of practitioners, as the findings are limited by its context~\cite{devanbu2016belief}.
As Shepperd notes, ``It would be highly unusual for a single primary study to provide a definitive answer to any question of much significance to practitioners''~\cite{shepperd2013combining}.

Consequently, many scholars advocate that ``we need to combine results from multiple primary studies''~\cite{shepperd2013combining} and that the field of SE research ``needs to move to a portfolio of empirical studies (on a single research hypothesis) being the norm rather than the currently unconvincing ‘one-off’, normally laboratory-based, studies that currently dominate the research literature''~\cite{miller2005replicating} and still do~\cite{shepperd2018role}.
These portfolios of replications (when replicating a controlled experiment also called \emph{families of experiments}~\cite{basili1999building}) strengthen the validity of research findings and generate more reliable conclusions~\cite{nosek2020replication}.

While portfolios of replications strengthen the knowledge creation phase of knowledge translation~\cite{santos2020research} by increasing the validity of drawn conclusions, they do not necessarily serve the knowledge application phase.
Indeed, portfolios of replications on their own even complicate maintaining an overview of all relevant primary studies, and contradicting results are non-trivial to reconcile~\cite{rosenthal2001meta}.
This necessitates the aggregation and integration of results from primary studies~\cite{ciolkowski2005accumulation}, commonly referred to as \emph{research synthesis}.
Research synthesis is a ``collective term for a family of methods that are used to summarize, integrate, combine, and compare the findings''~\cite{cruzes2011research} of individual pieces of evidence about the same phenomenon.
Shepperd summarized the synthesis process in five steps~\cite{shepperd2013combining}.

\begin{enumerate}
    \item \textbf{Problem formulation}: specifying a research question, usually about the influence of one independent variable on one dependent variable
    \item \textbf{Locating evidence}: searching literature that contains evidence about the research question
    \item \textbf{Appraising evidence quality}: applying quality inclusion criteria
    \item \textbf{Evidence synthesis and interpretation}: extracting relevant data and performing the research synthesis that infers about the initial research question based on the accumulated evidence
    \item \textbf{Reporting}: disseminating the results in a research report
\end{enumerate}

To implement step four, SE research has adopted several research synthesis methods from more mature empirical research disciplines~\cite{dixon2005synthesising,pickard1998combining,shepperd2013combining,dieste2011comparative,santos2020research,santos2019procedure}.
Among the most popular is the \emph{narrative synthesis}~\cite{santos2018analyzing}, a textual summary of research findings that is often criticized for its unsystematic approach~\cite{ciolkowski2009we} and lack of a quantitative outcome (i.e., a quantification of the effect strength of the observed phenomenon).
The more systematic \emph{vote-counting} classifies the effect of an independent variable on a dependent variable on ordinal scales, e.g., depending on its sign (i.e., positive, neutral, or negative effect)~\cite{pickard1998combining} or its strength of evidence (e.g., third party claims, circumstantial evidence, and strong evidence)~\cite{wohlin2013evidence}.
A histogram of classified findings from primary studies indicates the tendency of the effect observed by a portfolio of replications.
Similarly, the \emph{structured synthesis method} (SSM) proposed by dos Santos and Travassos~\cite{dos2013representation,dos2017structured} aggregates evidence from heterogeneous sources based on the belief value of these primary studies.

Empirical researchers usually consider \emph{meta-analysis} as the state-of-the-art for synthesizing quantitative results from controlled experiments~\cite{rosenthal2001meta}.
Other than vote-counting or the SSM, meta-analysis solely integrates reported statistics without introducing any researcher subjectivity beyond the selection of primary studies.
Various forms of meta-analysis exist and are adopted in SE research~\cite{santos2018analyzing}.
The two approaches considered superior are \emph{aggregate data} (AD) meta-analysis and \emph{stratified independent participant data} (IPD-S) meta-analysis~\cite{santos2018analyzing}.
AD meta-analysis pools the calculated effect sizes of primary studies together and calculates an overall effect size of the independent on the dependent variable~\cite{rosenthal2001meta}, often represented in a Forest plot~\cite{lewis2001forest}.
IPD-S meta-analysis pools together the raw data from all experiments of the primary studies and analyzes this data directly~\cite{riley2010meta}, but retains information about the belonging of each data point to the experiment it originated from to account for between-study variance~\cite{santos2018analyzing}.
IPD-S meta-analysis is commonly regarded as the gold standard for research synthesis~\cite{sutton2008recent} but is rarely applied in SE research~\cite{santos2018analyzing} since it depends on the availability of the raw data from every included primary study.
Among the different methods, narrative synthesis and AD meta-analysis remain most present in SE research~\cite{santos2018analyzing}.
Yet, meta-analysis remains uncommon in SE in general~\cite{kitchenham2020meta}.
Some examples from our field of research include synthesizing evidence about defect prediction~\cite{hosseini2017systematic,zain2023application} and TDD~\cite{rafique2012effects}.

\subsection{Theory Building}
\label{sec:related:theory}

Postulating and validating theories is a goal of any scientific discipline~\cite{fernandez2019empirical} to abstract from passing trends and provide practitioners with reliable, valid decision support~\cite{hannay2007systematic}.
SE mainly considers three types of theories~\cite{ralph2018toward}:

\begin{enumerate}
    \item \textbf{Theories for understanding} organize entities into meaningful categories.
    \item \textbf{Process theories} explain how something is happening.
    \item \textbf{Variance theories} quantify the strength of the effect of an independent variable on a dependent variable.
\end{enumerate}

Each type of theory serves a distinct purpose.
Variance theories quantify the effect strength of a phenomenon, which offers decision support to practitioners, e.g., when deciding whether to adopt a technique like TDD~\cite{rafique2012effects}.
To achieve this quantification, researchers need to synthesize individual pieces of quantitative evidence about a phenomenon.

\subsection{Validity of Variance Theories}
\label{sec:related:validity}

A SE practitioner could consult either a variance theory or a single piece of quantitative evidence for decision support.
The appeal of a variance theory over a single piece of evidence is not just the summary of relevant knowledge, but also that a variance theory aims to be more \emph{valid} than a single piece of evidence~\cite{hannay2007systematic}.
Because variance theories---just like controlled experiments---aim to quantify the effect strength of one or more independent variables on a dependent variable, the four dimensions of validity proposed by Cook and Campbell~\cite{cook1979quasi} and popularized in SE by Wohlin et al.~\cite{wohlin2012experimentation} apply similarly.
In the following subsections, we elaborate how variance theories aim to exceed individual studies in each dimension of validity.
We close each subsection with current limitations that motivate advancing the frontier of research synthesis.

\subsubsection{External Validity}
\label{sec:related:validity:external}

Most notably, a variance theory exceeds a single piece of evidence in \emph{external validity}.
External validity describes the degree to which a drawn conclusion can be generalized from the studied to the larger, often industrial context~\cite{wohlin2012experimentation}.
Since a variance theory synthesizes evidence from more than one study, it can cover a context larger than any individual study.
As such, the effect concluded by a variance theory is more likely to generalize to the target population~\cite{gomez2014understanding}.

However, producing replications eligible for synthesis is non-trival as their configuration can vary across several dimensions, e.g., operationalization, population, protocol, and experimenters~\cite{gomez2014understanding}.
This creates a spectrum ranging from replications as close as possible to the original study (referred to as literal~\cite{gomez2014understanding} or strict replications~\cite{juristo2012replication}) over replications varying one dimension (referred to as operational~\cite{gomez2014understanding} or differentiated replications~\cite{juristo2012replication}), to replications varying all dimensions and only keeping the underlying hypothesis, i.e., conceptual replications~\cite{gomez2014understanding,juristo2012replication}.
How the degree of variation between studies strengthens or threatens the external validity of a synthesized variance theory has been subject to extensive debate~\cite{kitchenham2008role,juristo2011role}.
Gomez et al. consolidated the debate with the proposal of \emph{systematic replications}, which structure multiple replications varying different dimensions to purposefully address potentially influencing factors~\cite{gomez2014understanding}.
But even these systematic replications could only detect \emph{if} any dimension influences the phenomenon of interest but not \emph{how}.
In such a case, a systematic replication would have identified a relevant effect but remains unable to quantify its significance.

\subsubsection{Internal Validity}
\label{sec:related:validity:internal}

A variance theory can also be more \emph{internally valid} than single pieces of evidence.
Internal validity describes the degree to which the drawn conclusion is causal and free of confounding effects~\cite{wohlin2012experimentation}.
Researchers usually achieve this by revising an existing hypothesis via adding or changing its causal assumptions.
For example, in the field of software testing, Inozemtseva and Reid revised the common claim that the coverage of a test suite correlates with its effectiveness by postulating that the \emph{test suite size} confounds this effect~\cite{inozemtseva2014coverage}.

This populates the literature about a specific phenomenon with evidence of varying internal validity.
In principle, research synthesis can maximize the internal validity of a variance theory by constraining the evidence location phase (cf. \Cref{sec:related:synthesis}) to only those pieces of evidence with maximum internal validity.
However, current research synthesis methods are not fit to act on these variations.
This is rooted in their focus on relationships between exactly two variables~\cite{miller2000applying,santos2018analyzing}.
Yet, it is unrealistic to assume that contemporary phenomena are isolated to only two variables, as human~\cite{pickard1998combining} and other context factors~\cite{pfleeger1999albert} usually interact with a phenomenon of interest.

Attempts to bend meta-analysis to deal with more complex causal assumptions beyond two-variable relationships produced subgroup or stratified analyses.
For example, Harris et al. clustered a synthesis study with 136 primary studies into 31 meta-analyses to isolate the effect of mediators~\cite{harris1985mediation}.
Similarly, the aforementioned meta-analysis by Rafique and Mi{\v{s}}i{\'c} involving 27 primary studies allowed a subgroup analysis revealing the mediating effect of the study population~\cite{rafique2012effects}.
However, all of these approaches are post-hoc attempts that insufficiently compensate the lack of an upfront, causal understanding of complex hypotheses.
Despite their additional level of complexity, they still fail to capture moderating or interaction effects~\cite{frattini2026towards}.% and categorically exclude quantitative evidence from observational studies~\cite{sutton2001bayesian} by design.

\subsubsection{Conclusion Validity}
\label{sec:related:validity:conclusion}

Variance theories can also improve the \emph{conclusion validity} of a claim over a single piece of evidence.
Conclusion validity describes the degree to which statistical assumptions are met when applying inferential statistics to derive a quantitative conclusion from data~\cite{wohlin2012experimentation}.
When the raw collected data is available---e.g., in IDP-S meta-analyses---this data can be analyzed with the best-fitting statistical method, regardless of what method has been applied to the individual pieces of evidence.
Such improvements to conclusion validity are typically driven by adopting more robust statistical methods~\cite{mcelreath2018statistical}, a method often referred to as a \emph{reanalysis} or \emph{test for robustness}~\cite{nosek2020replication}.
For example, Furia et al. reanalyzed two empirical studies about the effectiveness of automatically generated tests and programming languages performance with Bayesian statistics~\cite{furia2019bayesian}.
Compared to the baseline frequentist statistics, the Bayesian re-analysis violates fewer statistical assumptions and produces more meaningful insights~\cite{mcelreath2018statistical}.

While this example demonstrates the improvement of conclusion validity for a single piece of evidence, the same principle translates to variance theories.
However, optimizing the conclusion validity of a variance theory requires to systematically compare methods of statistical inference and selecting the one that fits the properties of the data best.
At the current point in time, research synthesis offers no advice on selecting an optimal statistical inference method.
Quite the opposite, as some synthesis techniques (like AD meta-analysis) are even locked into a frequentist statistical framework.

\subsubsection{Construct Validity}
\label{sec:related:validity:construct}

The last dimension of validity typically discussed in experimentation research is \emph{construct validity}.
Construct validity describes the degree to which measurements and instruments reflect a construct of interest, i.e., a value that cannot be directly measured~\cite{cook1979design}.
Construct validity is the only dimension of validity where a variance theory does not exceed, but is rather limited by the individual pieces of evidence.
In fact, varying construct validity of individual pieces of evidence even questions their synthesizability into one variance theory.
For example, consider two studies about the impact of TDD on developer productivity.
If the first study operationalizes productivity as ``number of lines of code written in a time frame'' and the second as ``self-reported assessment of productivity'', the dependent variables of the two pieces of evidence differ so vastly that a synthesis of the two effect strengths into one variance theory would not be meaningful from a quantitative perspective.\footnote{Qualitatively speaking, this form of triangulation can be valuable, but not for the quantification of an effect strength.}

\section{Goal and Method}
\label{sec:goal}

%\subsection{Goal}

Variance theories produced via research synthesis have the potential to be of greater internal, external, and conclusion validity than individual pieces of evidence.
Yet, as discussed in \Cref{sec:related:validity}, several shortcomings to the state-of-the-art in research synthesis limit this potential.
In this work, we propose a framework that systematically improves the external, internal, and conclusion validity of a variance theory about a phenomenon.
As such, it serves as a tool for research communities to orchestrate effective and constructive collaboration across individual studies.

%\subsection{Method}

Given the declared goal and the availability of partial methods (like meta-analysis), \textit{method engineering}~\cite{brinkkemper1996method}---in particular: method assembly~\cite{brinkkemper1998assembly}---becomes a viable research strategy.
Method engineering in SE was introduced by Brinkkemper as ``the discipline to construct new methods from parts of existing methods, called method fragments''~\cite{brinkkemper1998assembly}.
The three remediable threats to validity of variance theories elaborated in \Cref{sec:related:validity} can be individually addressed with the following, already existing method fragments~\cite{brinkkemper1998assembly}:

\begin{itemize}
    \item The threat to \textbf{external validity} can be addressed through replication studies~\cite{juristo2012replication,gomez2014understanding,nosek2020replication} and meta-analysis~\cite{yang2002meta,miller2000applying,santos2020research}.
    \item The threat to \textbf{internal validity} can be addressed via statistical causal inference~\cite{siebert2023applications,pearl2010causal}, particularly model comparison~\cite{mcelreath2018statistical,furia2022applying,cinelli2024crash}.
    \item The threat to \textbf{conclusion validity} can be addressed via statistical reanalysis~\cite{nosek2020replication,frattini2024second}.
\end{itemize}

As discussed in \Cref{sec:related:validity:construct}, given the absence of a proven method fragment to address the threat to construct validity, we do not include it in our framework. 
Efforts to remedy threats to construct validity~\cite{ralph2018construct} are commendable but remain heavily debated~\cite{sjoberg2022construct}.
To the best of our knowledge, there is no reliable method for systematically assessing the construct validity of an operationalization, and hence, no method to involve in our process theory.
The implications of this are minimal for variance theories.
It only implies that these variance theories quantify the relationships between \emph{operationalizations}, not \emph{concepts}.
While some scholars claim that ``theories are formulated at the conceptual level''~\cite{sjoberg2022construct}, we neither found evidence for this claim nor see any reason for it.
While theories on a conceptual level certainly expand their scope, this elevation comes at the cost of construct validity.
And as long as there is no reliable method for assessing construct validity, we decide this tradeoff in favor of validity over scope.
Consequently, we consider a valid conclusion about an effect between operationalizations no less valuable than between concepts.
Still, we discuss the full implication of this limitation and potential ways to overcome it in \Cref{sec:discussion:limitations}.

% categorization of methods according to Brinkkemper's dimensions
The method assembled from the three eligible method fragments is on a conceptual level of abstraction and on a method level of granularity~\cite{brinkkemper1998assembly}.
We specify the method from a process perspective that models the relationships and order of involved stages and activities.
This serves as a guide to a researcher conducting synthesis how to conduct the process.

\section{Framework for Research Synthesis}
\label{sec:framework}

Our framework for managing variance theories consists of a definition of evidence in \Cref{sec:framework:evidence} and, based on it, a framework of relationships between two pieces of evidence in \Cref{sec:framework:relationship}.
\Cref{sec:framework:evolution} then describes how these allow describing the synthesis and evolution of a variance theory.
%\Cref{sec:framework:impact} describes the implications of the framework on research synthesis practices in SE.

\subsection{Evidence}
\label{sec:framework:evidence}

To describe the relationship between evidence, we first need to formalize the concept of evidence.
We define a piece of empirical, quantitative evidence $e$ as a tuple $e:=E(h, d, m)$ consisting of three components.

\begin{itemize}
    \item \textbf{Hypothesis} $h$: A hypothesis consisting of variables and (assumed) causal relationships between those variables. For example, $h_1:=x \rightarrow y$ defines the hypothesis $h_1$ as variable $x$ causally influencing variable $y$.
    \item \textbf{Data} $d$: A record of observations of all variables contained in $h$. An eligible dataset $d_1$ for $h_1$ requires observation for both variables $x$ and $y$.
    \item \textbf{Method} $m$: An analysis method that processes the data $d$ under the hypothesis $h$ to produce a conclusion.
\end{itemize}

\begin{figure}
    \centering
    \includegraphics[width=0.85\linewidth]{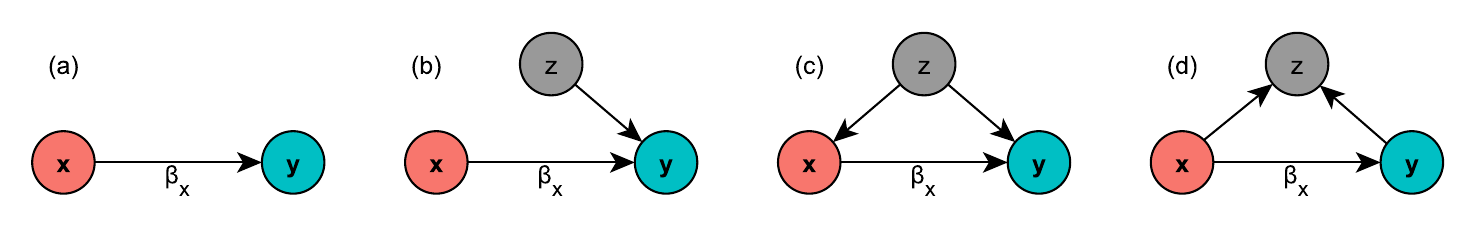}
    \caption{DAGs representing a hypothesis and three revisions}
    \label{fig:dags}
\end{figure}

Hypotheses are networks of variables and relationships among them.
As such, they can be visualized via directed, acyclic graphs (DAGs)~\cite{elwert2013graphical} as shown in \Cref{fig:dags}.
In these DAGs, nodes represent variables, and directed edges represent assumed causal relationships.
\Cref{fig:dags} (a) shows a graphical representation of a simple, two-variable hypothesis $h_1$.
More often, though, manuscripts present such simple hypotheses textually.
This often takes the form of a verbose null hypothesis, e.g., ``$x$ has no effect on $y$'' or ``There is no significant difference in values of $y$ for different values of $x$.''
More complex hypotheses involving more variables and relationships like \Cref{fig:dags} (c) and (d) require graphical representation but are rare in SE research~\cite{siebert2023applications}.

The two colored nodes in the four DAGs represent the main \textit{phenomenon} of interest, i.e., the independent variable or \emph{exposure} (colored red) and the \emph{outcome} or response variable (colored cyan).
The goal of a piece of empirical, quantitative evidence is to estimate the average causal effect (ACE) of the main exposure(s) on the dependent outcome variable.
Additional variables (colored grey) may be relevant to the hypothesis but not part of the main phenomenon under study.
For example, \Cref{fig:dags} (b) implies that $z$ also affects $y$, i.e., $z \rightarrow y$, in addition to $x$.
Therefore, the same phenomenon of interest can be described in multiple different hypotheses.

All analysis methods $m$ require deriving a \emph{statistical model} from the causal model---i.e., the hypothesis $h$~\cite{mcelreath2018statistical}.
A statistical model typically consists of a regression model, i.e., a specified, often linear relationship between one or more predictors and the outcome variable.
In the case of simple, two-variable hypotheses, this boils down to regressing the outcome on the only predictor.
For example, the statistical model derived from $h_1$ in \Cref{fig:dags} (a) would be $y \sim x$.
In the case of more complex hypotheses, one must select a subset of independent variables to condition on, called the \emph{adjustment set}.
This subset should maximize the precision of the estimation of the effect of $x$ on $y$ and, on the other hand, ensure that the causal effect is not confounded~\cite{cinelli2024crash}.
For example, the statistical model derived from \Cref{fig:dags} (c) would be $y \sim x + z$ where $z$ is included to de-confound the effect of $x$ on $y$.
The statistical model of \Cref{fig:dags} (d) would be $y \sim x$ since including $z$---called a \emph{collider}---would confound $x$ on $y$~\cite{mcelreath2018statistical}. 
%The subset that de-confounds the causal effect is commonly called the \emph{adjustment set}~\cite{cinelli2024crash} and can be determined by applying a systematic procedure called the \emph{backdoor adjustment}~\cite{pearl2010causal}.
%Explaining this procedure in detail goes beyond the scope of this manuscript but is well-explained in existing literature~\cite{mcelreath2018statistical,cinelli2024crash,pearl2010causal}.

The appropriateness of analysis methods depends on the complexity of the hypothesis and the properties of the variables.
For simple, two-variable hypotheses consisting of one independent and one dependent variable, most scholars resort to null hypothesis significance tests (NHSTs) like the Student's t-test or its variants~\cite{wagner2020challenges}.
Here, the choice depends on whether the dependent variable seems to follow a normal distribution and whether the data is paired or not.
For more complex hypotheses involving more than one independent variable, scholars tend to apply linear regression models with multiple predictors~\cite{vegas2015crossover}.

Applying the analysis method $m$ to the data set $d$ based on the causal model implied by the hypothesis $h$ produces the piece of evidence $e$ that offers a conclusion.
The nature of the conclusion depends on the analysis method.
For example, NHSTs propose a p-value that scholars commonly compare with an arbitrary significance level $\alpha$ to determine whether the independent variable evokes a statistically significant difference in the dependent variable.
For linear regression models, the conclusion takes the form of coefficients (e.g., $\beta_x$ in \Cref{fig:dags} (a), representing the strength of the impact of $x$ on $y$).
From these coefficients, one can additionally calculate confidence intervals for each independent variable.
If the confidence interval of a variable is not consistent with 0, i.e., it does not intersect 0, then the variable can be considered to have a significant impact on the dependent variable.

\subsection{Relationship between Evidence}
\label{sec:framework:relationship}

\begin{figure*}
    \centering
    \includegraphics[width=\linewidth]{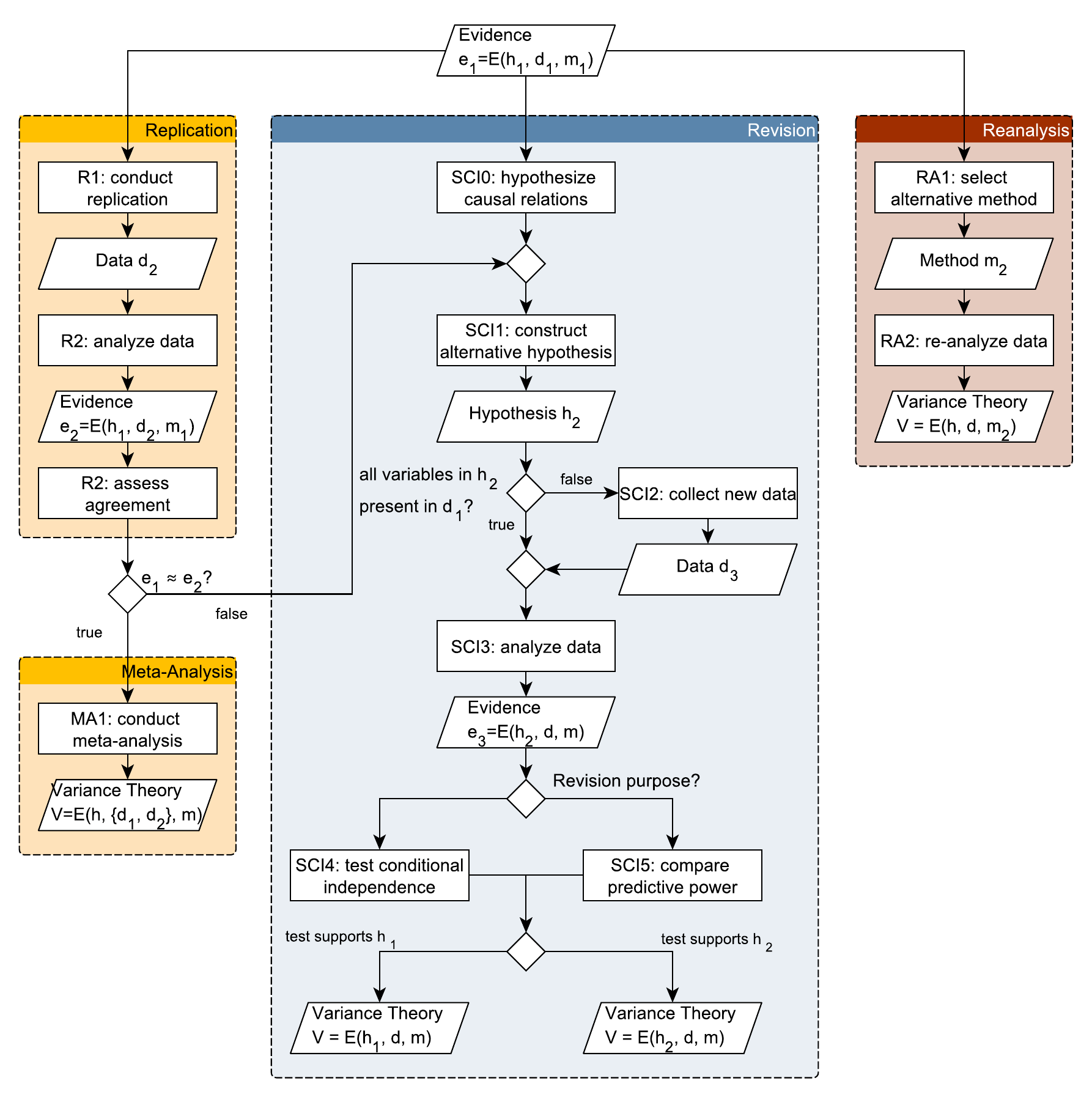}
    \caption{Method for research synthesis}
    \label{fig:framework}
\end{figure*}

Variance theories emerge from the synthesis of multiple pieces of empirical, quantitative evidence, which increases their validity and abstracts from passing trends~\cite{hannay2007systematic}.
To accommodate research synthesis that goes beyond the meta-analysis of a homogeneous set of primary studies~\cite{pickard1998combining}, the relationship between two pieces of evidence needs to be clear.
\Cref{fig:framework} visualizes process of synthesizing two or more pieces of evidence, which allows showing the evolution of evidence over time.
Starting from an initial piece of evidence $e_1=E(h_1, d_1, m_1)$, we consider three types visualized as paths in \Cref{fig:framework}.
We explain each type by referring to the three branches and the steps within them (to which we refer by their code, e.g., R1 for ``conducting a replication'').

\subsubsection{Replication}
\label{sec:framework:relationship:replication}

The most commonly known relationship between two pieces of empirical evidence in SE research is via \textit{replication} (left branch colored yellow in \Cref{fig:framework}).
%A replication is a type of study that offers diagnostic evidence about a previous empirical study~\cite{nosek2020replication}.
A replication subscribes to the same causal hypothesis $h_1$ and uses the same analysis method $m_1$ but collects a different data set $d_2$ to produce a new piece of evidence $e_2:=E(h_1, d_2, m_1)$ (R1).
The conclusion derived from the replication $e_2$ is compared with the conclusion of the original piece of evidence $e_1$ to check for agreement (R2).
Checking for agreement depends on the nature of the conclusion that the analysis method $m_1$ produces.
If $m_1$ is a type of hypothesis test that produces a p-value, this check is referred to as an \textit{aggregation of p-values}~\cite{santos2018analyzing} via Fisher's or Stouffer's method~\cite{borenstein2021introduction}.
If $m_1$ is a type of regression model that produces confidence intervals of coefficients, then the check boils down to assessing whether the confidence intervals overlap in a Forest plot~\cite{lewis2001forest}, or AD or IPD-S meta-analysis techniques~\cite{santos2018analyzing}.
Still, determining the agreement between replications is non-trivial~\cite{destefanis2026treplication}.

If the conclusions agree, the two pieces of evidence $e_1$ and $e_2$ can be synthesized to a variance theory $V=E(h_1, \{d_1, d_2\}, m_1)$.
This variance theory has an improved external validity over $e_1$ as the replication shows that the conclusion also in both contexts $d_1$ and $d_2$.
%Multiple replications can be synthesized using meta-analysis techniques---e.g., the aforementioned AD meta-analysis---to obtain an average confidence interval from the original study and all replications.
%If the conclusions agree, AD meta-analysis produces a more reliable conclusion about the strength of the causal effect of the phenomenon under study.
%The extent of improvement depends on how close the replication was.
%The more design elements of the original experiment were changed in the replication~\cite{gomez2010replications}, the greater the improvement of the external validity~\cite{juristo2011role}.
%For example, if the replication was performed at a different \textit{site} but the confidence intervals agree, then the replication concludes that the proposed causal relationship (i.e., that $x$ has a positive impact on $y$) is independent of the site.
However, in case the conclusions disagree, SE literature offers little advice on how to reconcile these results.
The disagreeing conclusions indicate a \emph{causal leak}~\cite{martinez2024decomposing}, i.e., that at least one variable that would explain the difference between $e_1$ and $e_2$ is missing from $h_1$.
A causal leak requires a \emph{revision} of the original hypothesis.

\subsubsection{Revision}
\label{sec:framework:relationship:revision}

A new piece of evidence acts as a \emph{revision} of another when the hypothesis $h_1$ is revised to a competing hypothesis $h_2$ (middle branch colored blue in \Cref{fig:framework}).
This hypothesis supposedly explains the phenomenon under study---which produced the data $d_1$ and $d_2$---better and, therefore, has greater internal validity.
The competing network can include new or discard existing variables or can---alternatively or additionally---propose different causal relationships between variables.
Only the variables belonging to the phenomenon under study (i.e., the main exposure and outcome) need to remain included.
Otherwise, the new hypothesis pertains to a different phenomenon.
\Cref{fig:dags} (b)--(d) visualize revisions of \Cref{fig:dags} (a) as they contain the variables of the main phenomenon under study ($x$ and $y$) but include an additional variable ($z$) and different relationships.

Disagreeing conclusions from a replication but also emerging qualitative evidence (SCI0) may trigger a revision.
For example, a qualitative study might suggest that the variable $z$ also influences $y$ even before observing disagreeing conclusions from replications.
Proposing a new hypothesis $h_2$ may also require collecting a new data set $d_3$ if $h_2$ contains variables not recorded in $d_1$ (SCI2).
In the abstract example, a new data set $d_3$ that records both $z$ in addition to $x$ and $y$ is necessary. 

With the new piece of evidence $e_3$ produced under the proposed hypothesis $h_2$ (SCI3), the competing hypotheses are evaluated by model comparison.
The type of comparison depends on the purpose of the revision.
Revisions can serve two different purposes~\cite{cinelli2024crash,navarro2019between}: increasing the precision of the effect estimation, and deconfounding the effect estimation.
The first purpose is to increase the precision of estimating the effect $\beta_x$ of $x$ on $y$.
For example, involving an additional, independent variable $z$ with an assumed causal relationship $z \rightarrow y$ (see \Cref{fig:dags} (b) visualizing such an alternative hypothesis $h_2$) may increase the precision of the estimate of the ACE~\cite{cinelli2024crash}.
By including $z$ in the statistical model, some of the variance previously attributed to $\beta_x$ will be explained by the influence of $z$, increasing the precision of the estimation of $\beta_x$.
The second possible purpose of a revision is to deconfound the estimation of the effect $\beta_x$ of $x$ on $y$~\cite{mcelreath2018statistical}.
This is particularly relevant to phenomena studied in observational, not experimental, settings where the exposure is not a true ``independent'' variable but can be influenced by factors other than random assignment.
A confounder could be a common cause as visualized in \Cref{fig:dags} (c) where variable $z$ impacts both $x$ and $y$, therefore biasing the direct effect of $x$ on $y$.
The adjustment set of this hypothesis includes $z$ as a predictor of $y$ to de-confound the effect of $x$ on $y$~\cite{mcelreath2018statistical}.

To evaluate a revision aiming at de-confounding, testable implications in the form of \emph{independencies} and \emph{conditional independencies} are derived from the hypotheses~\cite{mcelreath2018statistical} (SCI4).
For example, according to $h_3$ in \Cref{fig:dags} (c), both $x$ and $y$ depend on $z$. %, commonly noted as $x \not\Perp z$ and $y \not\Perp y$.
Conversely, $h_1$ in \Cref{fig:dags} (a) does not include this claim and implies that $x$ and $y$ are independent of $z$. %, i.e., $x \Perp z$ and $y \Perp y$.
Additionally, $h_3$ implies that the strength of the ACE of $x$ on $y$ changes when conditioning on $z$ via deconfounding, which $h_1$ does not imply.
%$y$ should be independent from $x$ once conditioning on $z$, often formalized as $(y \not\Perp x~|~z)$.
%Conditional independence can be determined from the data.
Correlational analyses on the available data set $d_3$ can confirm or refute these assumed independencies~\cite{mcelreath2018statistical}.
Comparing the statistical model $y \sim x$ (derived from $h_1$) with $y \sim x + z$ (derived from $h_3$) produces two estimates of the ACE of $x$ on $y$.
If the ACE is the same, then the effect of $x$ on $y$ is independent of $z$ and $h_1$ represents the causal relations of the phenomenon under study better.
If the ACE is different, then the effect depends on $z$, and $h_3$ is more valid.
Consequently, the internal validity of $h_3$ exceeds the one of $h_1$ and can be considered the currently superior causal model to explain the phenomenon under investigation.

To evaluate a revision aiming at increasing the precision, the out-of-sample predictive power of the two models is compared (SCI5) via an appropriate information criterion~\cite{mcelreath2018statistical}.
Metrics like the Akaike Information Criterion (AIC)~\cite{akaike2011akaike} or leave-one-out cross-validation (LOO) can be applied depending on the analysis method~\cite{vehtari2017practical,magnusson2020leave}.
These metrics assign scores to competing hypotheses and infer which predicts the observed data best.
The model with the greatest predictive power is assumed to be more internally valid.

While both purposes of revisions aim to strengthen the internal validity of a hypothesis, the respective evaluation that decides the comparison is not interchangeable.
Model comparison used to determine the hypothesis with the greater out-of-sample predictive power is not fit when aiming to deconfound evidence, as confounded evidence may very well exhibit a greater predictive power than a deconfounded one~\cite{mcelreath2018statistical}.
Consequently, the distinction of purpose when conducting a revision is imperative for the choice of evaluation method.

At every point in time, the hypothesis that shows the greatest internal validity should be the one that future studies should subscribe to.
This means that all future studies investigating the phenomenon of $x$ and $y$ should (1) record all variables involved in the hypothesis that are part of its adjustment set, and (2) analyze the collected data under the assumption of the internally most valid hypothesis.
The resulting variance theory optimizes the internal validity of the claim about the phenomenon.

Some scholars in SE research would consider revisions a type of replication, but only provide very approximate guidelines to reconcile the difference in hypotheses during synthesis~\cite{gomez2014understanding}.
Changing some elements of the experimental design in a revision is sometimes called a reproduction, differentiating it from a similar replication~\cite{cartwright1991replicability,juristo2012replication}.
To clearly distinguish the purpose of revisions from replications and emphasize that their decision process is dissimilar, we use the term ``revision'' instead of ``replication.''

\subsubsection{Reanalysis}
\label{sec:framework:relationship:reanalysis}

The least commonly discussed relationship between pieces empirical evidence in SE research is the \emph{reanalysis} of existing data (right branch colored red in \Cref{fig:framework}).
Reanalysis---sometimes also referred to as a ``test for robustness''~\cite{nosek2020replication}---describes the application of a different analysis method $m_2$ to the same data $d$ under the same causal assumptions $h$~\cite{gomez2010replications}.
In special cases, however, reanalysis may also be necessitated by a revision.
For example, extending a hypothesis to include two instead of one independent variable will make analysis methods that only operate with one independent variable (e.g., a t-test) ineligible and necessitate more complex ones (e.g., a linear model).

Reanalyses are mostly driven by adapting more advanced methods from other disciplines (e.g., statistics, or evidence-based medical research) (RA1).
One instance of this type of evolution is the ongoing endeavor to abandon simple NHSTs for more advanced Bayesian data analysis~\cite{furia2019bayesian}.
Reanalyses increase the conclusion validity of the evidence by revising statistical assumptions~\cite{wohlin2012experimentation}.
The decision of which analysis method to prefer over another is often based on intricate statistical comparisons~\cite{mcelreath2018statistical}, which SE researchers usually adapt and do not conduct themselves.
Re-analyzing existing data sets under given causal hypotheses (RA2) produces a variance theory with greater conclusion validity.

\subsection{Evolution}
\label{sec:framework:evolution}

The definition of a piece of evidence in \Cref{sec:framework:evidence} and how two pieces can relate to each other in \Cref{sec:framework:relationship} allow modeling the incremental evolution of research about a phenomenon.
This yields an \emph{evidence evolution graph} (EEG) which we illustrate in \Cref{tab:eeg} and describe as an example in this section.
As a starting point, an original study provides a piece of evidence $e_1$ (written $e1$ in the EEG for better readability) by applying the analysis method $m_1$ (e.g., a linear model) to data set $d_1$ under the causal assumption $h_1$ (defined as $x \rightarrow y$).
 
\begin{table}[!ht]
    \centering
    \caption{Exemplary evidence evolution graph}
    \label{tab:eeg}
    \begin{tabular}[b]{clllll}
        \toprule
        \textbf{EEG} & \textbf{Evolution Type}& \textbf{Hypothesis} & \textbf{Data} & \textbf{Analysis} & \textbf{Conclusion} \\
        \midrule
        \multirow[c]{6}{*}{ \includegraphics[height=72.5px]{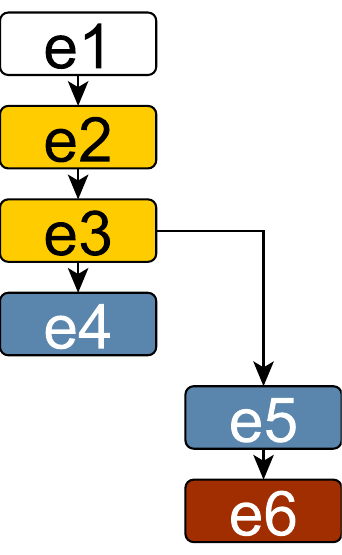}} 
        & (Original study) & $h_1: x \rightarrow y$ & $d_1$ & $m_1$ & $\text{CI}_x=[0.2, 0.6]$ \\
        & Replication & $h_1$ & $d_2$  & $m_1$ & \color{goodgreen} $\text{CI}_x=[0.15, 0.7]$ \\
        & Replication & $h_1$ & $d_3$  & $m_1$ & \color{badred} $\text{CI}_x=[-0.4, 0.2]$\\
        & Revision & $h_2: x \rightarrow y; x \leftarrow z \rightarrow y$ & $d_4$  & $m_1$ & \color{badred} {$z$ has no effect} \\
        & Revision & $h_3: x \rightarrow y; x \leftarrow a \rightarrow y$ & $d_5$  & $m_1$ & \color{goodgreen} {$a$ has an effect}\\
        & Reanalysis & $h_3$ & $d_5$  & $m_2$ & $\text{CI}_x=[-0.2, 0.2]$ \\
        \bottomrule
    \end{tabular}
  \end{table}

All subsequent pieces of evidence $e_n$ investigate the same phenomenon $x \rightarrow y$ but with specific relationships to the original studies.
$e_2$ and $e_3$ are replications of $e_1$, i.e., they subscribe to the same hypothesis $h_1$ and use the same analysis method $m_1$ but collect different data sets $d_2$ and $d_3$.
However, while the conclusion drawn from $e_2$ largely agrees with $e_1$ on a positive effect ($[0.2, 0.6] \approx [0.15, 0.7]$), $e_3$ disagrees ($[0.2, 0.6] \not\approx [-0.4, 0.2]$).
The disagreement of conclusions necessitates a revision.
$e_4$ revises the previous hypothesis by proposing $h_2$, according to which the variable $z$ confounds the effect $x \rightarrow y$.
However, if the conclusion shows that $z$ has no effect, i.e., it does not explain the differences that set $e_3$ apart from $e_1$ and $e_2$, then this revision does not have the intended effect.
$h_2$ has no greater internal validity than $h_1$ and is, hence, not worth pursuing further, constituting a ``dead end'' of the evidence evolution.
However, another hypothesis $h_3$ postulating that instead variable $a$ acts as the confounder might explain the differences previously observed, elevating it to the hypothesis with the greatest internal validity.
Future replications should subscribe to $h_3$ when contributing evidence to the phenomenon $x \rightarrow y$ and make sure to always collect data about $a$ as well to control the confounding.
Finally, a reanalysis $e_6$ might keep both $h_3$ and $d_5$ but use a different analysis method $m_2$ (e.g., a generalized linear model instead of a linear model) if the outcome variable $y$ is shown to fit a different distribution better.

%\section{Platform}
%\label{sec:platform}

\section{Application}
\label{sec:application}

To demonstrate the use of the proposed framework, we apply it to two active areas of SE research: test coverage (\Cref{sec:application:coverage}) and requirements quality (\Cref{sec:application:quality}).
Both areas implicitly aim at producing variance theories, i.e., quantifying how one factor influences another.
The application shows how the use of the framework can relate individual pieces of evidence to a holistic perspective, and simultaneously reveal issues of research synthesis in the absence of a framework.

\subsection{Test Coverage affecting Test Suite Effectiveness}
\label{sec:application:coverage}

\begin{figure}
    \centering
    \includegraphics[width=0.85\linewidth]{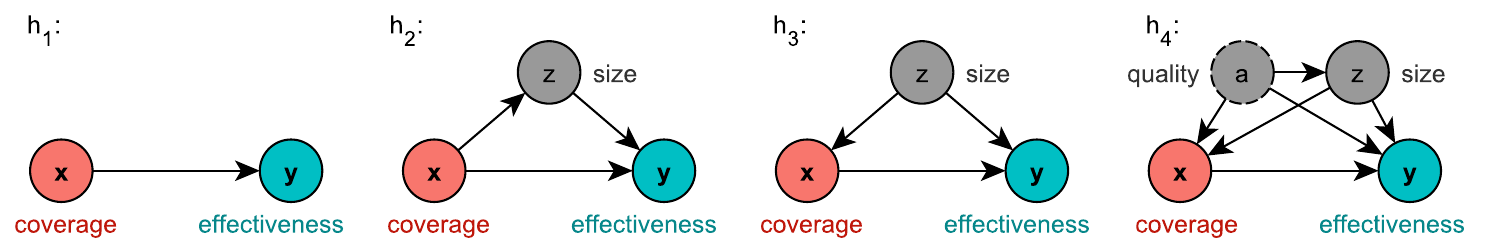}
    \caption{Hypotheses about the impact of coverage on test suite effectiveness}
    \label{fig:application:coverage:dags}
\end{figure}

Determining the effectiveness of a test suite, i.e., its fault detection capability, has been a long standing goal of the testing community.
Because the actual fault detection capability can only be determined in hindsight, the community proposed test adequacy criteria like \emph{coverage} metrics to predict the effectiveness of a test suite~\cite{chen2020revisiting}.
Goodenough and Gerhart postulated that adequacy criteria like coverage are meaningful proxies for test reliability~\cite{goodenough1975toward}.
This sparked a long empirical investigation of the claim.
\Cref{tab:application:coverage} applies the previously introduced framework to a selection of notable contributions to this line of research.

\begin{table}[!ht]
    \centering
    \caption{Evidence evolution about the effect of coverage on mutation scores}
    \label{tab:application:coverage}
    \begin{tabular}[b]{cllllll}
        \toprule
        \textbf{EEG} & \textbf{Type} & \textbf{Ref.} & \textbf{Hyp.} & \textbf{Data} & \textbf{Analysis} & \textbf{Conclusion} \\
        \midrule
        \multirow[c]{7}{*}{ \includegraphics[height=82px]{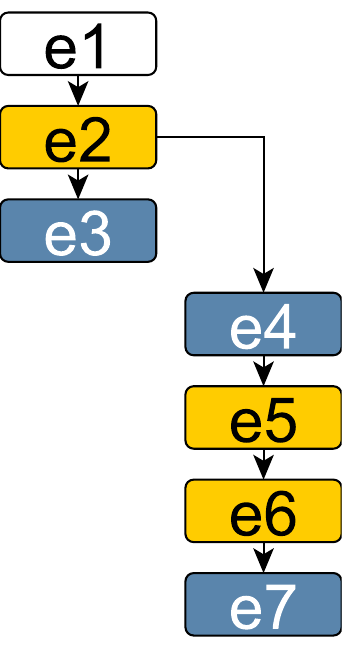}} 
        & (Original) & \cite{frankl1993experimental} & $h_1$ & 9 programs & Regression & \color{badred} Inconclusive\\
        & Replication & \cite{andrews2006using} & $h_1$ & \texttt{space.c} & Regression & Positive correlation \\
        & Revision & \cite{andrews2006using} & $h_2$ & \texttt{space.c} & Regression & Positive correlation \\
        & Revision & \cite{namin2009influence} & $h_3$ & 7 programs & Regression & Inconclusive \\
        & Replication & \cite{inozemtseva2014coverage} & $h_3$ & 5 Java programs & Visualization & \color{goodgreen} \texttt{size} acts as a confounder \\
        & Replication & \cite{gopinath2014code} & $h_3$ & 250 Java projects & Regression & \color{badred} \texttt{size} does not confound\\
        & Revision & \cite{chen2020revisiting} & $h_4$ & - & - & - \\
        \bottomrule
    \end{tabular}
  \end{table}

Frankl and Weiss~\cite{frankl1993experimental} tested the hypothesis that a test adequacy criterion like coverage affects test suite effectiveness, denoted as $h_1$ in \Cref{tab:application:coverage} and visualized in \Cref{fig:application:coverage:dags}.
Applying logistic regression to 9 programs ($e_1$), only 4 of these programs showed a correlation between coverage and effectiveness in fault-detection, rendering the piece of evidence inconclusive.
Andrews et al.~\cite{andrews2006using} replicated the experiment on the industrial-scale \texttt{space.c} data set but arrived at the conclusion $e_2$ that coverage does have a positive effect on effectiveness.
To reconcile the contradiction between $e_1$ and $e_2$, the authors additionally conjecture that the \emph{size} of a test suite may mediate the effect: 
test suites with greater coverage are also likely to be larger in size (i.e., contain more test cases), which could also improve the effectiveness.
Based on this revised hypothesis (see $h_2$ in \Cref{fig:application:coverage:dags}), the authors produced evidence $e_3$ supporting that both coverage and size have an effect on effectiveness.
However, Namin and Andrews~\cite{namin2009influence} later revise that claim again.
In the absence of a framework for causal inference, they claim that it is equally likely that size instead acts as a confounder (see $h_3$ in \Cref{fig:application:coverage:dags}).
However, when comparing $h_3$ to $h_1$, the authors do so by evaluating the model fit via the $R^2$ score of the two regression models.
As elaborated in \Cref{sec:framework:relationship:revision}, a revision for the purpose of de-biasing the effect of the phenomenon must be evaluated via testing conditional independencies, not model fit, because the model fit of a biased model could very well exceed the model fit of a de-biased one~\cite{mcelreath2018statistical}.
In other words: the statistical model derived from $h_1$ which does \emph{not} control for the confounder size suggested by $h_3$ may very well have an equal or better model fit.
Hence, Namin and Andrews~\cite{namin2009influence} arrive at an inconclusive piece of evidence where both hypotheses seem equally internally valid.
However, Inozemtseva and Holmes~\cite{inozemtseva2014coverage} later replicate this revision but with the correct data analysis method:
In their evidence $e_5$, they compare $h_1$ with $h_3$ on 5 Java programs by visualizing the association between \emph{coverage} and \emph{effectiveness}, once controlling \emph{size} ($h_3$) and once not ($h_1$).
The results clearly show that once controlling \emph{size} the association between \emph{coverage} and \emph{effectiveness} diminishes, supporting $h_3$ in that \emph{size} acts as a confounder.

Yet, when Gopinath et al.~\cite{gopinath2014code} replicated the study on 250 Java projects, they came to opposite conclusions in $e_6$:
When controlling \emph{size}, the association between \emph{coverage} and \emph{effectiveness} persisted.
To reconcile the once again contradicting replications, the study of Chen et al.~\cite{chen2020revisiting}---which included authors from both camps~\cite{inozemtseva2014coverage,gopinath2014code}---propose another revision (see $h_4$ in \Cref{fig:application:coverage:dags}):
While \emph{size} may indeed confound the effect of \emph{coverage} on \emph{effectiveness}, there may be yet another factor at play: 
the overall \emph{quality} of the test cases, which the authors explain via ``developer's desire to write effective tests''~\cite{chen2020revisiting}.
This overall quality may result both in more test cases (\emph{quality} $\rightarrow$ \emph{size}), greater coverage (\emph{quality} $\rightarrow$ \emph{coverage}), and more effective test cases (\emph{quality} $\rightarrow$ \emph{effectiveness}).
However, because the factor \emph{quality} was unobserved (indicated by the dashed lines around variable $a$ in $h_4$ in \Cref{fig:application:coverage:dags}), controlling \emph{size} did not sufficiently de-bias the effect of the phenomenon of interest, which potentially explains the contradictory results in $e_5$ and $e_6$.
Sadly, the authors did not provide empirical evidence for $h_4$.
However, they conclude with the recommendation ``that future studies should explicitly state their causal models and assumed underlying processes, which forces a clear statement of scientific questions and enables reasoning about whether the proposed experiments and analyses can answer that question.''~\cite{chen2020revisiting}.
This need is met by embedding causal modeling in the overall research synthesis framework.

\subsection{Passive Voice affecting Domain Model Completeness}
\label{sec:application:quality}

The application of the proposed framework to the field of requirements quality allows for even more nuanced insights.
Since authors of the manuscript at hand were also involved this research agenda~\cite{frattini2024applying,frattini2024second}, we can demonstrate how the application of the framework reveals several issues with research synthesis.
We show both the evolution of evidence as published (\Cref{sec:application:quality:published}) and how it \emph{should} have been conducted (\Cref{sec:application:quality:supposed}).

The research is centered around the question whether the use of \emph{passive voice} when specifying a requirement impacts the \emph{completeness of domain models} resulting from it~\cite{femmer2014impact}.
For example, when transforming the requirement ``All transaction details shall \emph{be obtained} from the Statement Database.'' into a domain model---i.e., a boxes-and-lines diagram of entities and associations described in the requirement---it is unclear who is responsible for ``obtaining transaction details.''
The active formulation ``\emph{The backend system shall obtain} all transaction details from the Statement Database.'' instead makes the association explicit.
Variance theories about the impact of passive voice on the completeness of domain models (e.g., ``Using passive voice increases makes incomplete domain models 45\% more likely.'') would help practitioners decide whether it is worth addressing and removing the hypothesized quality defect~\cite{frattini2023requirements}.

\subsubsection{Evolution as Published}
\label{sec:application:quality:published}

\begin{figure}
    \centering
    \includegraphics[width=0.7\linewidth]{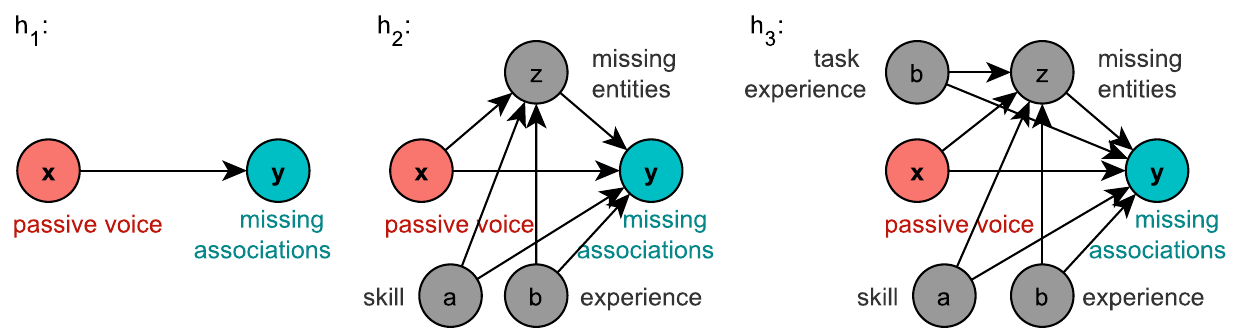}
    \caption{Hypotheses (as published) about the impact of passive voice on domain model completeness}
    \label{fig:application:passive:dags}
\end{figure}

Femmer et al.~\cite{femmer2014impact} contributed the first piece of empirical evidence $e_1$ about the impact of passive voice on domain model completeness.
The latter construct was measured by counting the number of missing actors, objects, and associations resulting from requirements written in either active or passive voice.
In the following, we focus on the effect on \emph{missing associations} (see $h_1$ in \Cref{fig:application:passive:dags}) and how evidence about it evolved (see \Cref{tab:application:passive}).

\begin{table*}[!ht]
    \centering
    \footnotesize
    \caption{Evolution of the variance theory on the impact of passive voice on domain modeling}
    \label{tab:application:passive}
    \begin{tabular}[b]{lllllll}
        \toprule
        \textbf{EEG} & \textbf{Evolution Type} & \textbf{Ref.} & \textbf{Hyp.} & \textbf{Data} & \textbf{Analysis Method} & \textbf{Conclusion} \\
        \midrule
        \multirow[c]{4}{*}{ \includegraphics[height=34px]{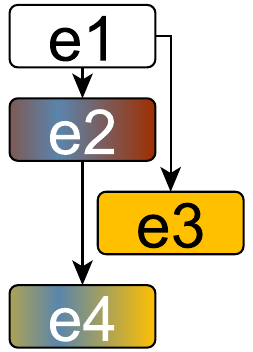}} & Original Study & 
        \cite{femmer2014impact} & $h_1$ & $d_1$: 15 students & $m_1$: Mann-Whitney U test & $p=0.001$ \\
        & Revision \& Reanalysis & \cite{frattini2024second} &  $h_2$ & $d_1$ & $m_2$: Bayesian model & $[-0.17, \sim0.49, +0.34]$ \\
        & Replication & \cite{frattini2024applying} &  $h_1$ & $d_2$: 25 participants & $m_1$: Wilcoxon signed-rank test & $p=0.025$ \\
        & Revision \& Replication & \cite{frattini2024applying} & $h_3$ & $d_2$ & $m_2$: Bayesian model & $[-0.25, \sim0.30, +0.45]$ \\
        \bottomrule
    \end{tabular}
\end{table*}

In a parallel-design controlled experiment with 15 university students generating domain models from 7 requirements each, the application of the Mann-Whitney U test resulted in a statistically significant p-value of $0.001$ and an effect size measured via Cliff's $\delta = 0.75$.
The authors concluded that the use of passive voice has a strong effect on the number of missing associations from domain models.
Frattini et al.~\cite{frattini2024second} revisited the claim by (1) revising the hypothesis to $h_2$ and (2) reanalysing the data with a Bayesian model instead of a frequentist significance test (cf. second row of \Cref{tab:application:passive}).
The latter was motivated by the improved conclusion validity of Bayesian models~\cite{furia2019bayesian,mcelreath2018statistical}.
The former was motivated by several conjectures:

\begin{enumerate}
    \item Domain models derived from passive-voice requirements may miss associations because they are missing entities (i.e., actors or objects) in the first place (i.e., \emph{passive voice} $\rightarrow$ \emph{missing entities} $\rightarrow$ \emph{missing associations}).
    \item The self-reported academic and industrial \emph{experience} may influence the number of missing elements.
    \item The subjects participating in the study may have different \emph{skill}, operationalized as a random effect in the analysis per participant ID.
\end{enumerate}

The resulting evidence $e_2$ retains only a very weak effect of passive voice: 
only in 34\% of all cases, passive voice causes more missing associations, while in 49\% they remain the same and in 17\% of all cases, passive voice sentences cause even fewer missing associations.

In a second follow-up study, Frattini et al.~\cite{frattini2024applying} contributed two pieces of evidence $e_3$ and $e_4$ (cf. third and forth row of \Cref{tab:application:passive}).
Based on a new data set $d_2$ collected from a within-subject design controlled experiment with 25 mostly industrial participants, they first replicated the original study.
Using the Wilcoxon signed-rank test (as the data in $d_2$ was paired, other than in $d_1$) under the assumption of $h_1$, evidence $e_3$ supported $e_1$ with $p=0.025$ (assuming a significance threshold of $\alpha = 0.05$).
Additionally, they conducted a joint revision and replication, i.e., $e_4$ further revises the hypothesis by challenging $h_2$ with $h_3$ (see \Cref{fig:application:passive:dags}) and uses $d_2$ instead of $d_1$.
The revised hypothesis was motivated $h_3$ by even more factors that potentially influence the outcome variable, like prior experience in domain modeling (labeled \emph{task experience} in \Cref{fig:application:passive:dags}).
$e_4$ agrees with $e_2$ in that the impact of passive voice on the number of missing associations from domain models is not strictly negative, as the likelihood of missing an association remains below 50\%.
However, $e_4$ suggests that the negative impact is more likely than assumed by $e_2$ (45\% instead of 34\%).

\subsubsection{Evolution as supposed to publish}
\label{sec:application:quality:supposed}

In the absence of an explicit framework to relate individual pieces of empirical evidence, the otherwise coherent research about the impact of passive voice on domain model completeness does not arrive at a proper variance theory.
Already the first follow-up study~\cite{frattini2024second} conflates a revision (i.e., changing the hypothesis from $h_1$ to $h_2$) with a reanalysis (i.e., changing the analysis method from a Mann-Whitney U test $m_1$ to a Bayesian model $m_2$), which makes it impossible to clearly attribute the diverging results to one of the changes.
Since we have access to all data and material, we were able to disentangle the implicit steps between $e_1$ and $e_2$, as outlined in \Cref{tab:application:passive:detail}.

\begin{table*}[!ht]
    \centering
    \caption{Decomposed evolution from $e_1$ to $e_2$}
    \label{tab:application:passive:detail}
    %\footnotesize
    \begin{tabular}[b]{llllll}
        \toprule
        \textbf{EEG} & \textbf{Evolution Type} & \textbf{Ref.} & \textbf{Hyp.} & \textbf{Analysis Method} & \textbf{Conclusion} \\
        \midrule
        \multirow[c]{8}{*}{ \includegraphics[height=100px]{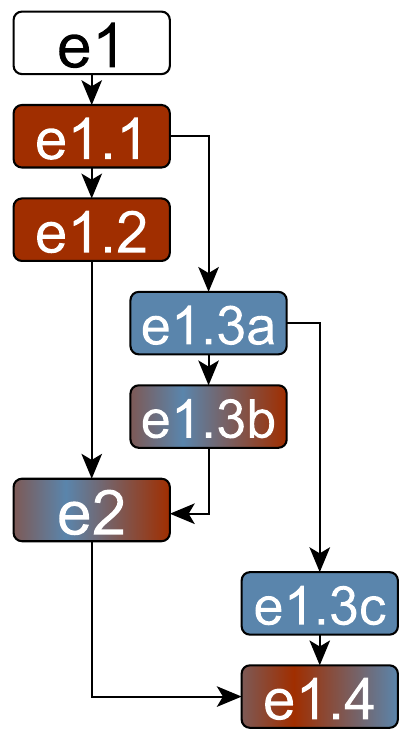}} & Original Study & \cite{femmer2014impact} & $h_1$ & $m_1$: Mann-Whitney U test & $p=0.001$ \\
        \hline
        & Reanalysis & & $h_1$ & $m_{1.1}$: linear model & $ci=[0.23, 0.84]$ \\
        & Reanalysis & & $h_1$ & $m_2$: Bayesian model & $[-0.15, \sim0.34, +0.51]$\\
        & Revision & & $h_{2a}$ & $m_{1.1}$: linear model & $ci=[0.17, 0.77]$ \\
        & Revision & & $h_{2}$ & $m_{1.2}$: linear mixed model & $ci=[-0.17, 0.82] $\\
        \hline
        & Revision/Reanalysis & \cite{frattini2024second} & $h_2$ & $m_2$: Bayesian model & $[-0.17, \sim0.49, +0.34]$ \\
        \hline
        & Revision & & $h_{2c}$ & $m_{1.2}$: linear mixed model & $ci=[0.03, 0.92]$ \\
        & Reanalysis/Revision & & $h_{2c}$ & $m_2$: Bayesian model & $[-0.14, \sim0.47, +0.39]$ \\
        \bottomrule
    \end{tabular}
\end{table*}

Resolving the conflation between the revision and reanalysis of $e_2$ in relation to $e_1$ prioritizes the reanalysis, since the revision requires an analysis method accepting more than one predictor (which the Mann-Whitney U test $m_1$ does not allow).
An intermediate reanalysis $e_{1.1}$ where the frequentist test is replaced by a standard linear regression model~\cite{lindelov2019common} eases the comparison with the Bayesian reanalysis.
The intermediate step shows that the pure reanalysis $e_{1.2}$ (i.e., $e_2$ without the revision but using the Bayesian model $m_2$) largely agrees with the frequentist test, although remaining more cautious. 
This contradicts the conclusion of the authors of the conflated revision and reanalysis, who attributed the observed difference to the use of Bayesian statistics~\cite{frattini2024second}.

\begin{figure}
    \centering
    \includegraphics[width=0.85\linewidth]{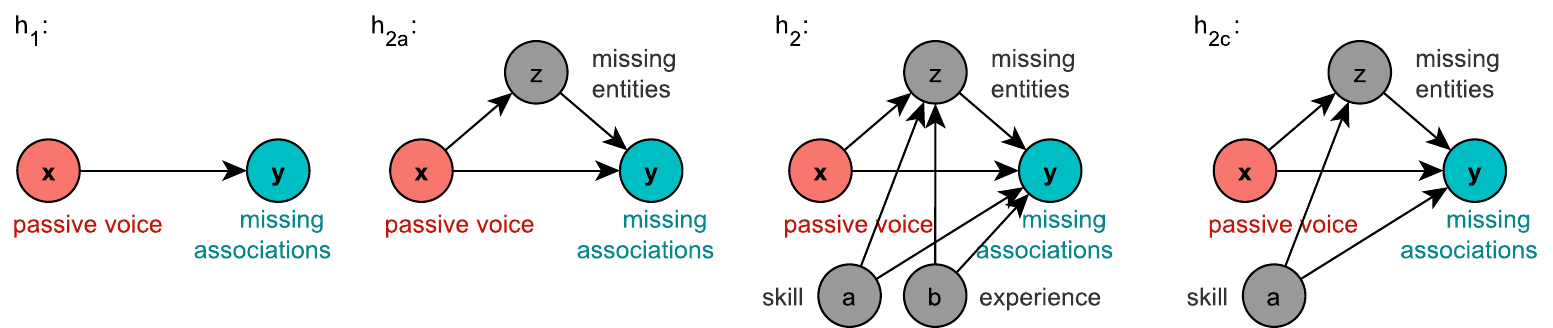}
    \caption{Hypotheses (as should be published) about the impact of passive voice on domain model completeness}
    \label{fig:dag:passive:zoomed}
\end{figure}

Replacing the simple significance test of $e_1$ with a linear model in $e_{1.1}$ allows systematically revising the hypothesis $h_1$ to isolate the impact of different additions.
The revision of $h_1$ to $h_2$ implies two revisions with two different purposes: 
the addition of the mediator \emph{missing entities} has causal implications~\cite{mcelreath2018statistical} while the addition of covariates like \emph{skill} serves the precision of the estimate~\cite{cinelli2024crash}.
Separating these two isolates their effect.
Adding only the mediator \emph{missing entities} ($h_{2a}$ in \Cref{fig:dag:passive:zoomed}) reduces the confidence interval from $[0.23, 0.84]$ to $[0.17, 0.77]$, but the conclusion remains the same.
Assessing the internal validity for this revision requires testing conditional independencies.
Including \emph{missing entities} as a mediator implies that \emph{missing associations} is not independent of it, which is shown by the authors: missing more entities causes missing more associations~\cite{frattini2024second}.
As such, $h_{2a}$ has greater internal validity than $h_1$.

However, once adding the covariate \emph{skill} operationalized as a subject specific random effect (which necessitates switching the linear model $m_{1.1}$ out for a linear \emph{mixed} model $m_{1.2}$) the confidence interval widens and overlaps with 0, such that the effect can no longer be considered statistically significant.
To determine whether the revision from $h_{2a}$ to $h_2$ has greater internal validity and whether the reanalysis from the linear model to the linear mixed model has greater conclusion validity, two checks are necessary.
Deciding greater conclusion validity requires assessing statistical properties of the two pieces of evidence.
For instance, the residuals of a linear model should be independent and identically distributed (iid)~\cite{west2022linear}.
Both the histograms of the residuals, the QQ-plot, and the Durbin-Watson test (autocorrelation greater than 0 ($p=0.07$)) suggest that the residuals of $e_{1.1}$ are not iid.
Consequently, applying a linear model $m_{1.1}$ may lead to invalid conclusions, and the conclusion validity of a linear mixed model $m_{1.2}$ is greater~\cite{west2022linear}.

Deciding greater internal validity requires assessing predictive power in this case.
We calculate the AIC, which applies to both $m_{1.1}$ and $m_{1.2}$~\cite{vaida2005conditional}, achieving scores of $AIC(e_{1.3a})=249.1$ and $AIC(e_{1.3b})=251.2$.
The score differential of about 2 points is considered negligible when interpreting the AIC values~\cite{greenwood2021intermediate}.
Hence, there is no strong evidence that $h_2$ is more internally valid than $h_{2a}$.

Finally arriving at the evidence $e_2$~\cite{frattini2024second} is now consistent from both angles:
$e_2$ acts as a revision of $e_{1.2}$ by changing only the hypothesis $h_1$ to $h_2$, for which the Bayesian model concludes that the effect of passive voice is smaller (i.e., causing more missed associations in only 34\% instead of 51\% of all cases).
Simultaneously, $e_2$ acts as a reanalysis of $e_{1.3b}$ by changing the linear mixed model to a Bayesian model.
$e_2$ concludes that in the majority of cases (49\%), the use of passive voice does not in- or decrease the number of missed associations from a domain model, which the confidence interval $[-0.17, 0.82]$ overestimates.

However, the decomposed evolution of evidence showed that there is little support for $h_2$ to be more internally valid than $h_1$.
Yet, the diagnostics of $e_{1.3b}$ reveal that the inclusion of \emph{skill} as a random effect had a strong impact:
While $e_{1.3b}$ only has a marginal $R^2$ value of 0.184, it has a conditional $R^2$ value of 0.561, which indicates that the random effects explain a lot more of the variance of the outcome variable than the fixed effects~\cite{nakagawa2013general}.
This is in line with observations in empirical research, that---when analyzing experiments with small sample sizes---the within-subject variance of subjects (e.g., their skill) has much more impact on the outcome than other measured variables~\cite{gelman2007data,baayen2008mixed}.
The benefit of the random effects in $h_2$ may be offset by the number of predictors, as the AIC metric penalizes an increased number of predictors to avoid overfitting~\cite{greenwood2021intermediate}.
This suggests the competing hypothesis $h_{2c}$ which retains the provenly effective random effects for participants' \emph{skill} but discards the fixed effects of academic and industrial \emph{experience}.
The operationalization of the fixed effect is questionable, as it was simply measured on an ordinal scale with four levels.
Therefore, the inclusion of these fixed effects might not benefit the estimation and rather overfit the estimation.
The predictive power $AIC(e_{1.3c}) = 241.4$ is significantly---i.e., more than 2 units~\cite{greenwood2021intermediate}---lower than $AIC(e_{1.3a})=249.1$ and $AIC(e_{1.3b})=251.2$, suggesting that it has the greatest internal validity and should have been used subsequently instead of $h_2$.
Evidence $e_{1.3c}$ concludes that passive voice has an impact of $ci_{e_{1.3c}}(passive)=[0.03, 0.92]$, which is still not consistent with 0 but broader than $ci_{e_{1.3b}}(passive)=[0.17, 0.77]$.
Using the same line of reasoning as before, the final piece of evidence $e_{1.4}$ could have combined the internally most valid hypothesis with analysis method of the highest conclusion validity, the Bayesian model $m_2$.
$e_{1.4}$ concludes that using passive voice in requirements causes fewer missed associations in 14\%, equal in 47\%, and more in 39\% of all cases.

\section{Discussion}
\label{sec:discussion}

As shown in the previous demonstrations, the framework proposed in \Cref{sec:framework} assembles method fragments to a coherent and consistent approach to improving the internal, external, and conclusion validity of variance theories.
\Cref{sec:discussion:implications} discusses the general implications of the framework, and \Cref{sec:discussion:usecases} elaborates three concrete use cases by which the framework can improve SE research.
However, we also acknowledge limitations in \Cref{sec:discussion:limitations} which necessitate the future work described in \Cref{sec:discussion:future}.

\subsection{Implications}
\label{sec:discussion:implications}

The proposed framework has the ability to impact both the knowledge creation, synthesis, and application phase of empirical research, effectively supporting the complete epistemological life cycle. 
Most prominently, the framework provides the terminology and process to synthesize individual pieces of empirical evidence about a phenomenon. 
This has the potential to improve the aggregation and integration of evidence in SE research and clearly identify how the frontier of the body of knowledge evolves.
At the same time, the framework also clearly values the contribution of replications, revisions, and reanalyses.
This increases the incentive to pursue follow-up research and not abandon empirical investigations after just one single study.

By extension, the synthesis process centered around evidence evolution graphs can also improve the knowledge creation phase.
Future studies contributing new evidence to an existing phenomenon will be able to better gauge the current frontier of a body of knowledge, and therefore, inform their design.
Proven hypotheses with maximum internal and analysis methods with maximum conclusion validity suggest which variables to collect and how to analyze data recording them.

This produces more coherent research that---after systematic synthesis---also becomes more digestible for practitioners in the form of focused, integrated results with maximum internal, external, and conclusion validity.
Such results have a higher chance of application in practice as they are more convincing than isolated one-off studies.
As a consequence, empirical SE research has the chance of contributing more practically relevant support for the intended target population of software engineers.

\subsection{Use Cases}
\label{sec:discussion:usecases}

Such implications can be realized via several use cases that improve the state of empirical SE research.
We present these use cases in ascending order of complexity, starting with the most accessible one.

\subsubsection{Framing new Evidence}

The framework helps researchers to position a piece of evidence in relation to an existing body of knowledge.
It provides a terminology to frame how new studies advance the body of knowledge with respect to existing ones:
New evidence can be classified as either a replication, revision, or reanalysis, depending on what parts of the empirical evidence changed in respect to its origin, and the framework encourages changing only one part at a time.
More importantly, the framework supports deciding whether this new evidence strengthens or challenges the body of knowledge:

\begin{enumerate}
    \item \textbf{Replications} (i.e., changing the data) where the conclusion 
    \begin{enumerate}
        \item agrees with the predecessor \emph{improve the external validity}.
        \item disagrees with the predecessor \emph{necessitate a revision}.
    \end{enumerate}
    \item \textbf{Revisions} (i.e., changing the hypothesis) \emph{improve the internal validity} if their purpose is
    \begin{enumerate}
        \item increasing the precision of the ACE estimation and the revision has a higher predictive power.
        \item deconfounding the ACE estimation and the revision is more consistent with (conditional) independencies.
    \end{enumerate}
    \item \textbf{Reanalyses} (i.e., changing the analysis method) \emph{improve conclusion validity} if the new analysis method violates fewer statistical assumptions.
\end{enumerate}

\subsubsection{Systematic Literature Reviews}

In extension to the prior use case, the framework allows to model the complete evolution of empirical evidence about a phenomenon, as shown in \Cref{sec:application}.
This aligns with the original purpose of a systematic literature review (SLR) as originally intended by Kitchenham et al.~\cite{kitchenham2007guidelines}.
By framing all empirical, quantitative studies investigating one phenomenon using the framework, the evolution of a variance theory about that phenomenon can be described, analyzed, and allows to make explicit and informed decision about future research directions.
Applying our framework to an SLR would produce a central EEG as shown in \Cref{sec:application}, which clearly shows which causal assumptions about the phenomenon are best supported (i.e., have the highest internal validity) and which are ``dead ends''.
This reveals the current empirical knowledge quantitative studies on a phenomenon and can inform future study design.

\subsubsection{Informing a follow-up Study}

Once an EEG of a phenomenon exists, researchers planning to conduct a follow-up study about the phenomenon can consult the ECG to improve the design of their study.
Two aspects revealed by an EEG are particularly relevant.
Firstly, the hypothesis with the highest internal validity informs which variables the researchers should collect.
For example, in the context of test coverage research (cf. \Cref{sec:application:coverage}), future research should at least collect data on \emph{test suite size} following Inozemtseva and Reid~\cite{inozemtseva2014coverage}.
Even better, additionally attempting to measure or control the underlying \emph{quality} aspiration with which the test suites were created could bring empirical evidence to the claim of Chen et al.~\cite{chen2020revisiting}.
Secondly, the analysis method with the highest conclusion validity informs how the researchers should analyze their data.
For example, in the context of requirements quality research (cf. \Cref{sec:application:quality}), Bayesian data analysis methods have shown to violate fewer statistical assumptions~\cite{frattini2024second}, and hence, should be preferred when conducting a new study.

\subsection{Limitations}
\label{sec:discussion:limitations}

Despite the potential advantages for empirical SE research, the proposal remains subject to several threats and limitations.

\subsubsection{Reliance on Fragments' Validity}
\label{sec:discussion:limitations:fragments}

Most prominently, the validity of the framework relies on the validity of the method fragments it is assembled from.
Since prior research extensively supports the validity of replications~\cite{nosek2020replication,gomez2010replications}, meta-analysis~\cite{yang2002meta,kitchenham2020meta}, statistical causal inference~\cite{pearl2010causal,mcelreath2018statistical}, and reanalysis~\cite{mcelreath2018statistical}, we deem this threat constrained.

\subsubsection{Limitation to quantitative Evidence}
\label{sec:discussion:limitations:quantitative}

By design of managing \emph{variance} theories, the proposed framework focuses on integrating quantitative evidence.
Qualitative evidence can trigger the revision of a hypothesis (cf. \Cref{sec:framework:relationship:revision}), but does not qualify as an individual piece of evidence involved in the synthesis due to its lack of quantification.

\subsubsection{Construct Validity}
\label{sec:discussion:limitations:construct}

The proposed framework circumvents construct validity entirely by focusing on variance theories about effects between \emph{operationalizations}, not concepts.
While this eliminates all threats to construct validity, it also limits the applicability of the resulting variance theories.
For example, the variance theory regarding test coverage in \Cref{sec:application:coverage} pertains to the effect of the operational \emph{test coverage}, not the broader concept of \emph{test adequacy criteria}.
Given the lack of a reliable method fragment to improve construct validity~\cite{sjoberg2022construct}, the framework contributes no systematic improvement along this dimension of validity.

\subsubsection{Quality Appraisal}
\label{sec:discussion:limitations:quality}

As with any form of secondary study, the quality appraisal of primary studies---i.e. the decision, which primary studies to include in the synthesis and which to exclude---is non-trivial~\cite{kitchenham2007guidelines}.
The proposed framework does not address this issue and only focuses on the relation and comparison of individual pieces of evidence once included.
The careful appraisal and selection of primary studies to integrate remains an ongoing challenge.

\subsection{Future Work and Roadmap}
\label{sec:discussion:future}

The retrospective application of the proposed framework has the potential to provide additional structure to SLRs.
Centering SLRs around evidence evolution graphs with explicit hypotheses, data, and analysis methods clearly describes the progress and frontier of research about a phenomenon.
For existing SLRs where this information is available, one avenue of future work is their replication using the proposed synthesis framework, similar to \Cref{sec:application:coverage,sec:application:quality}.
The main impact of the framework, however, is expected on new SLRs and new empirical evidence contributing to SLRs.
We envision a future of empirical SE research with the prospective adoption of the framework to not only synthesize quantitative evidence to variance theories, but also inform research design and dissemination.
Future work can further improve this impact.

From the perspective of researchers, the next step is facilitating the adoption of the framework.
We encourage researchers to structure their research efforts around a visual EEG that summarizes the evolution of empirical evidence hitherto.
Such EEG would act as a central guidance system coordinating future research about the same phenomenon.
A natural extension would be an online platform hosting the EEG where new evidence about the same phenomenon could be submitted, integrated, and maintained, similar to how Github acts as a platform enabling the distributed version control of source code repositories.
Even further, such platform could enable the automatic synthesis of all pieces of evidence following the same hypothesis and using the same analysis method.
This would mark a paradigm shift away from a static, incidental, and manual towards dynamic and continuous synthesis of primary studies.

From the perspective of practitioners, the next step is accessing the synthesized knowledge via these platforms.
If the variance theory about a phenomenon synthesized from all available and recent pieces of evidence is available to SE practitioners, SE research could improve the aspired knowledge translation~\cite{santos2020research} by presenting incrementally synthesized variance theories rather than unconvincing one-off studies to practitioners~\cite{miller2005replicating}.

\section{Conclusion}
\label{sec:conclusion}

Research synthesis can improve knowledge translation by producing variance theories with greater validity than individual pieces of evidence.
To extend synthesis practices in SE research beyond meta-analyses which only apply under specific conditions, we propose a definition of empirical, quantitative evidence and a framework for its evolution.
The framework allows to clearly relate two pieces of evidence about a phenomenon and supports deciding which has the greater internal, external, and conclusion validity.
Putting multiple pieces of evidence into relation with each other reveals the evolution of empirical research about a phenomenon and indicates its current frontier.
We believe that this vision for orchestrating empirical research about a phenomenon of interest will make research progress explicit, support the knowledge generation, synthesis, and application process, and enable productive, distributed research endeavors in our field of SE research.
We hope it will help us move beyond isolated islands of individual studies, and toward more coherent, constructive, and reliable research outcomes.

\section*{Conflict of Interest Statement}

The authors declare that the research was conducted in the absence of any commercial or financial relationships that could be construed as a potential conflict of interest.

\section*{Author Contributions}

The authors have contributed as follows according to the Contributor Role Taxonomy (CRediT):

\begin{itemize}
    \item \textbf{Julian Frattini}: conceptualization, data curation, investigation, methodology, project administration, validation, visualization, writing
    \item \textbf{Jannik Fischbach}: conceptualization, data curation, validation, writing
    \item \textbf{Davide Fucci}: conceptualization, investigation, validation, supervision, writing
    \item \textbf{Michael Unterkalmsteiner}: conceptualization, supervision, methodology, writing
    \item \textbf{Daniel Mendez}: conceptualization, supervision, methodology, writing
    \item \textbf{Robert Feldt}: conceptualization, supervision, validation, writing
    \item \textbf{Richard Torkar}: conceptualization, supervision, validation, writing
\end{itemize}

\section*{Funding}
This work was supported by the KKS foundation through the S.E.R.T. Research Profile project and the GIST project at Blekinge Institute of Technology.

\section*{Acknowledgments}

We express our deep gratitude to the participants of the ISERN workshop who critically reviewed this work and helped revise it.
Particularly, we thank Stefan Wagner, Sebastian Baltes, Maria Teres Baldassarre, Guilherme Horta Travassos, Jefferson Seide Molleri, Andreas Jedlitschka, Stefan Biffl, and Laurie Williams for their constructive feedback.

\section*{Supplemental Data}
No supplemental beyond the replication package is available.

\section*{Data Availability Statement}
All study data is publicly available in our replication package~\cite{replicationpackage}.

\bibliographystyle{materials/bibstyles/Frontiers-Vancouver}
\bibliography{materials/references}

\end{document}